\documentstyle [12pt] {article}

\def\be{\begin{equation}}
\def\ee{\end{equation}}
\def\ba{\begin{array}}
\def\ea{\end{array}}
\def\bea{\begin{eqnarray}}
\def\eea{\end{eqnarray}}
\def\GeV{{\rm GeV}}
\def\tr{{\rm tr}}
\def\Tr{{\rm Tr}}
\def\thefootnote{\fnsymbol{footnote}}
\def\chib{{\bar\chi}}
\def\psib{{\bar\psi}}
\def\nn{\nonumber}

\def\wS{S}
\def\wT{T}
\def\sS{{\cal S}}
\def\sT{{\cal T}}

\def\NPB#1#2#3{{Nucl.~Phys.} {\bf{B#1}} (19#2) #3}
\def\PLB#1#2#3{{Phys.~Lett.} {\bf{B#1}} (19#2) #3}
\def\PRD#1#2#3{{Phys.~Rev.} {\bf{D#1}} (19#2) #3}
\def\PRL#1#2#3{{Phys.~Rev.~Lett.} {\bf{#1}} (19#2) #3}

\if@twoside\oddsidemargin27mm
\else\oddsidemargin27mm\evensidemargin27mm\marginparwidth27mm\fi%
\begin{document}
\begin{titlepage}
{\sf
\begin{flushright}
{BONN--TH--97--10}\\
{SFB--375/295, TUM--HEP--311/98}\\
{TU--531, RCNS--97--05}\\
{January 1998}
\end{flushright}}
\vfill
\vspace{-1cm}
\begin{center}
{\large \bf Supersymmetry Breakdown at a Hidden Wall}

\vskip 1.2cm
{\sc H. P. Nilles$^{1,2}$, {\ }M. Olechowski$^{3,\ast}$ {\ and} 
{\ \sc M. Yamaguchi$^{4}$}\\
\vskip 1.5cm
{\em $^1$Physikalisches Institut, Universit\"at Bonn} \\
{\em Nussallee 12, D--53115 Bonn, Germany} \\
\vskip 1cm
{\em $^2$Max--Planck--Institut f\"ur Physik} \\
{\em D--80805 M\"{u}nchen, Germany}
\vskip 1cm
{\em $^3$Physik Department,Technische Universit\"at M\"unchen} \\
{\em D--85747 Garching, Germany}
\vskip 1cm
{\em $^4$Department of Physics, Tohoku University} \\
{\em Sendai 980--77, Japan}}
\end{center}
\vfill

\thispagestyle{empty}

\begin{abstract}

We consider hidden sector supersymmetry breakdown in the strongly
coupled heterotic $E_8\times E_8$ theory of Ho\v{r}ava and Witten.
Using effective field theory methods in four dimensions, we can
show that gravitational interactions induce soft breaking terms
in the observable sector that are of order of the gravitiono
mass. We apply these methods to the mechanism of gaugino
condensation at the hidden wall. Although the situation is
very similar to the weakly coupled case, there is a decisive
difference concerning the observable sector gaugino mass;
with desirable phenomenological as well as cosmological 
consequences.

\end{abstract}

\vskip 5mm \vskip0.5cm
\hrule width 5.cm \vskip 1.mm
\noindent
{\small\small ${}^{\ast}$ On leave of absence from 
Institute of Theoretical Physics, Warsaw University, Poland.\\}
\end{titlepage}

\def\thefootnote{\arabic{footnote}}
\setcounter{footnote}{0}
%
%
\section{Introduction}

{}From all the new and interesting results in string dualities,
it is the heterotic M--theory of Ho\v{r}ava and Witten
\cite{HW} that seems to have immediate impact on the discussion
of the phenomenological aspects of these theories.
One of the results concerns the question of the unification
of all fundamental coupling constants \cite{W} and the second
one the properties of the soft terms (especially the gaugino
masses) once supersymmetry is broken \cite{NOY}. In both cases results
that appeared problematic in the weakly
coupled case get modified in a satisfactory way, while the
overall qualitative picture remains essentially unchanged.

The heterotic M--theory is an 11--dimensional theory with the
$E_8\times E_8$ gauge fields living on two 10--dimensional
boundaries (walls), respectively, while the gravitational fields
can propagate in the bulk as well. A $d=4$ dimensional theory
with $N=1$ supersymmetry emerges at low energies when 6 dimensions 
are compactified on a Calabi--Yau manifold. The scales of that theory
are $M_{11}$, the $d=11$ Planck scale, $R_{11}$ the size of
the $x^{11}$ interval, and $V\sim R^6$ the volume of the
Calabi--Yau manifold. The quantities of interest in $d=4$,
the Planck mass, the GUT--scale and the unified gauge coupling
constant $\alpha_{GUT}$ should be determined through these
higher dimensional quantities. The fit of ref.\ \cite{W}
identifies $M_{GUT}\sim 3\cdot 10^{16}$ GeV  with the inverse 
Calabi--Yau radius $R^{-1}$. Adjusting $\alpha_{GUT}=1/25$ gives  
$M_{11}$ to be a few times larger than $M_{GUT}$. On the other hand, 
the fit of the actual value of the Planck scale can be achieved by
the choice of $R_{11}$ and, interestingly enough, $R_{11}$ turns out 
to be an order of magnitude larger than the fundamental 
length scale $M_{11}^{-1}$.
A satisfactory fit of the $d=4$ scales 
is thus possible, in contrast to the case of the weakly coupled
heterotic string, where the string scale seemed to be a factor
20 larger than $M_{GUT}$.

Otherwise the heterotic $E_8\times E_8$ string looks rather
attractive from the point of view of phenomenological
applications. One seems to be able to accommodate the correct
gauge group and particle spectrum. The mechanism of 
hidden sector gaugino
condensation leads to a breakdown of supersymmetry with
vanishing cosmological constant to leading order. With a
condensate scale $\Lambda\sim 10^{13}$ GeV, one obtains a
gravitino mass in the TeV range and soft scalar masses in that
range as well. In the simplest models \cite{DIN,DRSW,DIN2} this
type of supersymmetry breakdown is characterized through the vacuum
expectation value of moduli fields other than the dilaton, giving
a small problem with the soft gaugino masses in the observable
sector: they turn out to be too small, generically some two 
orders of magnitude smaller than the soft scalar masses. 
It is again in the framework of heterotic M--theory that 
this problem is solved \cite{NOY}; gaugino masses are of 
the same size as (or even larger than) the soft scalar masses.

The mechanism of hidden sector gaugino condensation itself
can be realized in a way very similar to the weakly
coupled case. This includes the mechanism of cancellation
of the vacuum energy, which in the weakly coupled case arises
because of a cancellation of the gaugino condensate with a
vacuum expectation value of the three index tensor field $H$ of
$d=10$ supergravity. This cancellation is at the origin of the
fact that supersymmetry breakdown is dominated by a $T$ modulus
field rather than the dilaton ($S$). Ho\v{r}ava \cite{H} observed that
this compensation of the vacuum expectation values of the condensate
and $H$ carries over to the M--theory case.
In \cite{NOY} we explicitly worked out the mechanism of
gaugino condensation in the heterotic M--theory  
and showed the similarity to the weakly coupled case. Now the gaugino
condensate forms at the hidden 4--dimensional wall and is
cancelled {\bf locally} at that wall by the vacuum expectation
value (vev) of a Chern--Simons term. This also clarifies some 
questions concerning the nature of the vev of $H$ that arose in
the weakly coupled case. 
 
In the present paper we want to discuss the phenomenological
properties of the heterotic M--theory. We also give details of the
calculation leading to those results that were presented
in \cite{NOY}. This includes a presentation of the full effective
four--dimensional $N=1$ supergravity action in leading and
next--to--leading order, the mechanism of hidden sector
gaugino condensation and its explicit consequences for
supersymmetry breaking and the scalar potential and finally the
resulting soft breaking terms in the 4--dimensional theory.
We will at each step first explain the situation in the weakly
coupled theory and then compare it to the results obtained in
the M--theory case.

These results are obtained using the method of reduction and
truncation that has been successfully applied to the weakly
coupled case \cite{W2}. It is a simplified prescription that shows the
main qualitative features of the effective $d=4$ effective theory.
In orbifold compactification it would represent the fields and
interactions in the untwisted sector.
We compute K\"ahler potential ($K$), superpotential ($W$) and gauge
kinetic function ($f$) both in the weakly and strongly coupled regime
and explain similarities and differences.

The results in leading order had been obtained previously
\cite{AQ,LLN,DG,D}. These papers mainly focused on the
breakdown of supersymmetry via a Scherk--Schwarz mechanism. It
remains to be seen, if and how such a mechanism can be
related to the mechanism of gaugino condensation.

After our paper \cite{NOY} appeared,
several other groups also considered a discussion of the
effective action beyond leading order as well as
the mechanism of gaugino condensation \cite{LT,LOW,CKM,LOW2}.

The paper will proceed as follows. In chapter 2 we discuss the scales 
and the question of unification as suggested in \cite{W}
and compare the two cases. In chapter 3 we review the old results
in the weakly coupled heterotic string, including important
corrections to the $f$--function at the one loop level.
Chapter 4 deals with the effective $d=4$ action of M--theory
using the method of reduction and truncation. In this case we have
to deal with a nontrivial obstruction first encountered in
\cite{W}. It leads to an explicit $x^{11}$ dependence 
of certain fields, which is induced by vevs of antisymmetric
tensor fields at the walls. To obtain the effective action
in $d=4$ we have to integrate out this dependence. This then leads to
corrections to $K$ and $f$ in next to leading order, which are very 
similar compared to those in the weakly coupled case. We also
discuss the appearance and the size of a critical radius for $R_{11}$.
The phenomenological fit presented in chapter 2 implies that we are 
not too far from that critical radius.
In chapter 5 we discuss the mechanism of gaugino condensation.
We start with the weakly coupled case and investigate the 
nature of the vev of the $H$--field (concerning some
quantization conditions) and the cancellation of the vacuum energy. 
We then move to the strongly coupled case and see that such 
a cancellation appears locally at one wall. This supports the 
interpretation that the gaugino condensate is matched by
a nontrivial vev of a Chern--Simons term. We then explicitly
identify the mechanism of supersymmetry breakdown and the
nature of the gravitino. The goldstino turns out to be the
fermionic component of the $T$ superfield that represents
essentially the radius of the 11th dimension. It is a bulk field,
with a vev of its auxiliary component on one wall. Integrating
out the 11th dimension we then obtain explicitly the mass of
the gravitino.
Chapter 6 then deals with the induced soft breaking terms in
the observable sector: scalar and gaugino masses. We point out
a strong model dependence of the scalar masses and argue that they
are not too different from the gravitino mass. This all is
very similar to the situation in the weakly coupled case.
We then compute the soft gaugino masses and see that in the strongly 
coupled case they are of the order of the gravitino mass.
This comes from the fact that we are quite close to the critical
radius and represents a decisive difference to the weakly 
coupled regime.
In chapter 7 we discuss some immediate phenomenological
consequences, give an outlook and mention some open questions.


%
%
\section{Scales and unification}

The framework of string theory might ultimately lead to an 
explanation of the unification of all fundamental coupling
constants. In contrast to usual grand unified models describing
exclusively gauge interactions we here have a unification
with the gravitational interaction as well. One therefore expects 
the grand unified scale $M_{GUT}$ to be connected to the Planck
scale.

%
%
\subsection{Weakly coupled $E_8\times E_8$ heterotic string}

Models of particle physics that are derived as the low energy
limit of the $E_8\times E_8$ heterotic string are particularly
attractive. They seem to be able to accommodate the correct
gauge group and particle spectrum to lead to the supersymmetric
extension to the $SU(3)\times SU(2)\times U(1)$ standard model.
It is exactly in this framework that a unification of the
gauge coupling constants is expected to appear at a scale
$M_{GUT}=3\cdot 10^{16}$ GeV. This heterotic string theory
(weakly coupled at the string scale) in fact gives a 
prediction for the relation between gauge and gravitational
coupling constants. To see this explicitly let us have a look 
at the low energy effective action of the $d=10$--dimensional
field theory:
\be
L=-{4\over (\alpha^{\prime})^3} \int
d^{10}x \sqrt{g} \exp({-2\phi})
\left(
{1\over (\alpha^{\prime})}R + {1\over 4}\tr F^2 + \ldots
\right),
\label{eq:10d}
\ee
where $\alpha^{\prime}$ is the string tension and $\phi$ the
dilaton field in $d=10$. A definite relation between gauge and
gravitational coupling appears because of the universal
behaviour of the dilaton term in eq.\ (\ref{eq:10d}). The 
effective $d=4$--dimensional theory is obtained after compactification
on a Calabi--Yau manifold with volume $V$:
\be
L=-{4\over (\alpha^{\prime})^3} \int
d^{4}x \sqrt{g} \exp({-2\phi}) V
\left(
{1\over (\alpha^{\prime})}R + {1\over 4}\tr F^2 + \ldots
\right).
\label{eq:4d}
\ee
Thus a universal factor $V\exp(-2\phi)$ multiplies both the $R$ and
$F^2$ terms. Newton's and Einstein's
gravitational coupling constants are related as
\be
G_N={1\over 8\pi}\kappa_4^2={1\over M_{\rm Planck}^2},
\label{eq:GN0}
\ee
with $M_{\rm Planck}\approx 1.2\cdot 10^{19}$ GeV. From
eq.\ (\ref{eq:4d}) we then deduce:
\be
G_N  
= {\exp(2\phi)(\alpha^{\prime})^4 \over 64\pi V }
\,,
\label{eq:GNW}
\ee
as well as
\be
\alpha_{GUT} = {\exp(2\phi)(\alpha^{\prime})^3 \over 16\pi V},
\label{eq:alphaGUTW}
\ee
leading to the relation
\be
G_N
={\alpha_{GUT}\alpha^{\prime}\over 4}.
\label{eq:GNR}
\ee
Putting in the value for $M_{\rm Planck}$ and 
$\alpha_{GUT}\approx 1/25$ one obtains a value for the string scale 
$M_{\rm string}=(\alpha^{\prime})^{-1/2}$ that is in the region of 
$10^{18}$ GeV. This is apparently much larger than the GUT--scale of
$3\cdot 10^{16}$ GeV, while naively one would like to identify
$M_{\rm string}$ with $M_{GUT}$. The discrepancy of the scales is 
sometimes called the unification problem in the framework of the 
weakly coupled heterotic string. Of course, the above argumentation 
is rather simple and more sophisticated (threshold) calculations are 
needed to settle this issue. In any case, the natural appearance of
$M_{\rm string}\sim M_{GUT}$ would have been desirable. 
Let us now see how the situation looks in the case of 
heterotic string theory at stronger coupling. 

%
%
\subsection{$E_8\times E_8$ M--theory}
\vskip5mm

The effective action of the strongly coupled
$E_8 \times E_8$ -- $M$--theory in the ``downstairs'' 
approach is given by
\cite{HW} 
(we take into account the numerical corrections found in 
\cite{CC})
\bea
L
\!\!&=&\!\!
{1\over \kappa^2} \int_{M^{11}}
d^{11}x \sqrt{g}
\left[
       - \frac{1}{2}R
       - \frac{1}{2} \psib_I \Gamma^{IJK} D_J
            \left( \frac{\Omega + {\hat\Omega}}{2} \right) \psi_K
       - \frac{1}{48} G_{IJKL} G^{IJKL}
\right.
\nn\\
&&\qquad\qquad\qquad\quad
       - \frac{\sqrt{2}}{384}
            \left( \psib_I \Gamma^{IJKLMN} \psi_N
                  +12 \psib^J \Gamma^{KL} \psi^M \right)
            \left( G_{JKLM} + {\hat G}_{JKLM} \right)
\nn\\
&&\qquad\qquad\qquad\quad
       - \left. \frac{\sqrt{2}}{3456}
            \epsilon^{I_1 I_2 \ldots I_{11}} C_{I_1 I_2 I_3}
            G_{I_4 \ldots I_7} G_{I_8 \ldots I_{11}}
\right]
\\
\!\!&+&\!\!
\frac{1}{4\pi(4\pi\kappa^2)^{2/3}}  \int_{M^{10}_i} 
d^{10}x \sqrt{g}
\left[ 
    - \frac{1}{4} F^a_{iAB} F_i^{aAB}
       - \frac{1}{2} \chib_i^a \Gamma^AD_A ({\hat\Omega}) \chi_i^a
\right.
\nn\\
&&\qquad\qquad\qquad\quad
\left.
       - \frac{1}{8} \psib_A \Gamma^{BC} \Gamma^A
            \left( F^a_{iBC} + {\hat F}^a_{iBC} \right) \chi_i^a
       + \frac{\sqrt{2}}{48}
            \left( \chib_i^a \Gamma^{ABC} \chi_i^a\right){\hat G}_{ABC11}
\right]
\nn
\eea
where $M^{11}$ is the ``downstairs'' manifold 
while $M_i^{10}$ are its 10--dimensional boundaries. 
In the lowest approximation $M^{11}$ is just
a product $M^4 \times X^6 \times S^1/Z_2$.
Compactifying to $d=4$ in such an approximation we obtain
\cite{W,CC}
\be
G_N 
= {\kappa_4^2\over 8 \pi} 
= {\kappa^2 \over 8 \pi R_{11} V}
\,,
\label{eq:GN}
\ee
\be
\alpha_{GUT} = {(4\pi\kappa^2)^{2/3} \over V}
\label{eq:alphaGUT}
\ee
with $V$ the volume of the Calabi--Yau manifold $X^6$
and $R_{11} = \pi\rho$ the $S^1/Z_2$ length.

The fundamental mass scale of the 11--dimensional theory is given by 
$M_{11} = \kappa^{-2/9}$. Let us see which value of $M_{11}$ is
favoured in a phenomenological application. For that purpose we
identify the Calabi--Yau volume $V$ with the GUT--scale: 
$V\sim(M_{GUT})^{-6}$. From (\ref{eq:alphaGUT}) and the value of
$\alpha_{GUT}=1/25$ at the grand unified scale, we can then deduce
the value of $M_{11}$
\be
V^{1/6} M_{11} 
=
(4\pi)^{1/9} \alpha_{GUT}^{-1/6}
\approx
2.3
\,,
\label{eq:VM11}
\ee
to be a few times larger than the GUT--scale. In a next step we
can now adjust the gravitational coupling constant by choosing
the appropriate value of $R_{11}$ using (\ref{eq:GN}). 
This leads to
\be
R_{11} M_{11}
=
\left(\frac{M_{Planck}}{M_{11}}\right)^2
\frac{\alpha_{GUT}}{8\pi(4\pi)^{2/3}}
\approx
2.9 \cdot 10^{-4} \left(\frac{M_{Planck}}{M_{11}}\right)^2
\,.
\label{eq:R11M11}
\ee
This simple analysis tells us the following: 

\begin{itemize}

\item
In contrast to the
weakly coupled case ( where we had a prediction (\ref{eq:GNR})),
the correct value of $M_{\rm Planck}$ can be fitted by adjusting
the value of $R_{11}$.

\item
The numerical value of $R_{11}^{-1}$ turns out to be 
approximately an order of magnitude smaller than $M_{11}$.

\item
Thus the 11th dimension appears to be larger than the dimensions
compactified on the Calabi--Yau manifold, and at an intermediate
stage the world appears 5--dimensional with two 4--dimensional
boundaries (walls).

\end{itemize}

We thus have the following picture of the evolution and unification
of coupling constants. 
At low energies the
world is 4--dimensional and the couplings evolve accordingly with
energy: a logarithmic variation of gauge coupling constants and
the usual power law behaviour for the gravitational coupling.
Around $R_{11}^{-1}$ we have an additional 5th dimension and the
power law evolution of the gravitational interactions changes.
Gauge couplings are not effected at that scale since the
gauge fields live on the walls and do not feel the existence of
the 5th dimension. Finally at $M_{GUT}$ the theory becomes
11--dimensional and both gravitational and gauge couplings
show a power law behaviour and meet at the scale $M_{11}$,
the fundamental scale of the theory. It is obvious that  
the correct choice of $R_{11}$ is needed to achieve unification.
We also see that, although the theory is weakly coupled at
$M_{GUT}$, this is no longer true at $M_{11}$. The naive
estimate for the evolution of the gauge coupling constants between
$M_{GUT}$ and $M_{11}$ goes with the sixth power of the scale.
At $M_{11}$ we thus expect unification of the couplings
at $\alpha\sim O(1)$. In that sense, the M--theoretic description
of the heterotic string gives an interpolation between
weak coupling and moderate coupling. In $d=4$ this is not
strong--weak coupling duality in the usual sense. We shall later
come back to these questions when we discuss the appearance of
a critical
limit on the size of $R_{11}$.

These are, of course, rather qualitative results. In order to
get a  more quantitative feeling for the range of $M_{11}$ and 
$R_{11}$, let us be a bit more specific and write 
the relation of the  
unification scale $M_{GUT}$ to the characteristic size 
of the Calabi--Yau space as:
\be
V^{1/6} = a M_{GUT}^{-1}
\,.
\label{eq:VMGUT}
\ee
In the spirit of our phenomenological ansatz we expect 
the parameter $a$ to be of order unity. 
Using the above identification and the value of 
$M_{GUT} = 3 \cdot 10^{16}$ GeV we obtain:
\be
M_{11} \approx \frac {2.3}{a} M_{GUT}
\,.
\ee
As said before, the scale $M_{11}$ occurs to 
be of the order of the unification 
scale $M_{GUT}$. However, we do not expect $M_{11}$ to be 
smaller than $M_{GUT}$ because we need the ordinary logarithmic 
evolution of the gauge coupling constants up to $M_{GUT}$. 
In fact, $M_{11}$ should be somewhat bigger in order to allow 
for the evolution of $\alpha$ from its unification value 1/25 
to the strong regime. 
Thus, we obtain phenomenologically interesting solutions if 
the parameter $a$ is quite close to 1. 
Putting the above value of $M_{11}$ into eq.\ (\ref{eq:R11M11}) 
we get the length of $S^1/Z_2$:
\be
R_{11} \approx 9.2 a^2 M_{11}^{-1} \approx 4 a^3 M_{GUT}^{-1}
\,.
\ee
It is about one order of magnitude bigger than the scale 
characteristic for the 11--dimensional theory. 
This is the reason for the relatively large value of the 
$d=4$ Planck Mass. Of course $R_{11}$ can not be too large. 
For $a < 2.3$ (values corresponding to $M_{11} > M_{GUT}$) 
we obtain $R_{11}^{-1} > 6.2 \cdot 10^{14}$ GeV 
(as we discussed, the parameter $a$ should not be too 
different from 1 which gives  $R_{11}^{-1}$ 
close to $7.4 \cdot 10^{15}$ GeV). 
Smaller values of $R_{11}^{-1}$ seem to be very unnatural. 
Trying to push $R_{11}^{-1}$ to smaller values would
need a redefinition of $M_{11}$. For that purpose in \cite{AQ}
a definition $m_{11}=2\pi(4\pi\kappa^2)^{-1/9}$ was used.
This allows then to push $a$ to as high values as $2\pi$.
With these rather extreme choices of both $a$ and $m_{11}$ one 
would then be able to obtain $R_{11}^{-1}$ as small as 
$3 \cdot 10^{13}$ GeV. Values smaller than that (like
values of $10^{12}$ GeV as sometimes quoted in the
literature) cannot be obtained. In any case, even values
in the lower $10^{13}$ GeV range seem to be in conflict with
the critical value of $R_{11}$, as we shall see in 
chapter 4.

%
%
\section{Weakly coupled heterotic string}

%
%
\subsection{The classical action}

We shall start from the $d=10$ effective field 
theory:
\be
L=-{4\over (\alpha^{\prime})^3} \int
d^{10}x \sqrt{g} \exp({-2\phi})
\left(
{1\over (\alpha^{\prime})}R + {1\over 4}\tr F^2 + {1\over 12}
\alpha^{\prime}H^2 + \ldots
\right),
\label{eq:10d+H}
\ee
where we have included the three index tensor field strength
\be
H=dB+\omega^{YM}-\omega^{L}.
\label{eq:Hfield}
\ee
$B$ is the two--index antisymmetric tensor while
\be
\omega^{YM}= \Tr(AF -{2\over 3}A^3)
\label{eq:omegaYM}
\ee
and
\be
\omega^{L}= \Tr(\omega R -{2\over 3}\omega^3)
\label{eq:omegaL}
\ee
are the Yang--Mills and Lorentz--Chern--Simons terms, respectively.
The addition of these terms in the definition of $H$ is 
needed for supersymmetry and anomaly freedom of the theory.

To obtain the effective theory in $d=4$ dimensions 
we use as an approximation the method of reduction and truncation
explained in ref.\ \cite{W2}. It essentially corresponds to
a torus compactification, while truncating states to arrive
at a $d=4$ theory with $N=1$ supersymmetry.
In string theory compactified on an
orbifold this would describe the dynamics of the untwisted sector. 
We retain the usual moduli fields $\wS$ and $\wT$ as well as matter
fields $C_i$ that transform nontrivially under the observable sector 
gauge group. In this approximation, the K\"ahler potential is given by 
\cite{W2,DIN2}
\be
G = - \log (\wS + \wS^*) - 3 \log (\wT + \wT^* - 2 C_i^* C_i) + 
\log \left| W \right|^2  
\label{eq:G}
\ee
with superpotential originating from the Chern--Simons terms
$\omega^{YM}$ \cite{DIN}
\be
 W(C) = d_{ijk} C_iC_jC_k   
\ee
and the gauge kinetic function is given by the dilaton field
\be
f = \wS \,. 
\label{eq:f}
\ee
For a detailed discussion of this method and the explicit
definition of the fields see the review
\cite{HPN}. These expressions for the $d=4$ effective action
look quite simple and it remains to be seen whether this
simplicity is true in general or whether it is an artifact of the
approximation. Our experience with supergravity models tells us
that the holomorphic functions $W$ and $f$ might be protected
by nonrenormalization theorems, while the K\"ahler potential is
strongly modified in perturbation theory. In addition we have to
be aware of the fact that the expressions given above are at best
representing a subsector of the theory. In orbifold compactification
this would be the untwisted sector, and we know that the
K\"ahler potential for twisted sectors fields will look quite
different. Nonetheless the used approximation turned out to be
useful for the discussion of those aspects of the theory that
determine the dynamics of the $T$-- and $S$--moduli. When trying to
extract, however, detailed masses and other properties of the fields 
one should be aware of the fact, that some results might not be
true in general and only appear as a result of the
simplicity of the approximation.

%
%
\subsection{One--loop corrections}

Not much can be said about the details of the corrections to the
K\"ahler potential. This has to be discussed on a model by model basis.
The situation with the superpotential is quite easy. There we expect
a nonrenormalization theorem to be at work. The inclusion of other 
sectors of the theory will lead to new terms in the superpotential
that in general have $T$--dependent coefficients. Such terms can be
computed in simple cases by using e.g. methods of conformal field
theory \cite{LMN}.

The situation for $f$, the gauge--kinetic function is more interesting.
Symmetries and holomorphicity lead us to believe, that although there
are nontrivial corrections at one--loop, no more perturbative
corrections are allowed at higher orders \cite{SV,N}. The existence 
of such corrections at one loop seems to be intimately connected to 
the mechanism of anomaly cancellation in the $d=10$ theory 
\cite{CK,IN}. To see this consider one of the anomaly cancellation 
counter--terms introduced by Green and Schwarz \cite{GS}:
\be
\epsilon^{VOLKSWAGEN}B_{VO} \Tr F^2_{LKSW} F^2_{AGEN}
\,.
\label{eq:GSterm}
\ee
We are interested in a $d=4$ theory with $N=1$ supersymmetry, 
and thus expect nontrivial vacuum expectation values for the 
curvature terms $\Tr R^2$ and field strengths $\Tr F^2$ in the 
extra six dimensions. Consistency of the theory requires  a 
condition for the 3--index tensor field strength. For $H$ to be
well defined, the quantity
\be
dH = \Tr F^2 - \Tr R^2
\label{eq:consistency}
\ee
has to vanish cohomologically \cite{Cetal}. 
In the simplest case (the so--called
standard embedding leading to gauge group $E_6\times E_8$) one chooses
equality pointwise $\Tr R^2 = \Tr F^2$. Let us now assume that
$\Tr F^2_{agen}$ is nonzero. The Green--Schwarz term given above
by eq.\ (\ref{eq:GSterm}) then leads to 
\be
\epsilon^{mn}B_{mn}\epsilon^{\mu\nu\rho\sigma}
\Tr F_{\mu\nu}F_{\rho\sigma}
\label{eq:FFtilde}
\ee
in the four--dimensional theory. An explicit inspection of the fields
tells us that $\epsilon^{mn}B_{mn}$ is the pseudoscalar axion that
belongs to the $T$--superfield. Upon supersymmetrization the term
in eq.\ (\ref{eq:FFtilde}) will then correspond to a one--loop
correction to the holomorphic $f$--function
 (\ref{eq:f}) that is proportional to $T$ with the coefficient
fixed entirely by the anomaly considerations. This is, of course,
nothing else than a threshold correction. In the simple case of the
standard embedding with gauge group $E_6\times E_8$ one obtains
e.g. 
\be
f_6 = \wS +\epsilon \wT\,;
\qquad\qquad
f_8 = \wS - \epsilon \wT\,.
\label{eq:f6f8}
\ee
respectively, where $\epsilon$ is the constant fixed by the
anomaly. These results can be backed up by explicit
calculations in string theory. In cases where such an
explicit calculation is feasible, many more details about these
corrections can be deduced.
The above result (\ref{eq:f6f8}) obtained in $d=10$
field theory represents an approximation of the exact result 
 in the large $T$--limit.  For a detailed discussion of 
these calculations and the limiting procedure see \cite{NS}. 
We have here
mainly concentrated on that limit, because it represents a rather
model independent statement.

Thus we have seen that there are corrections to the gauge--kinetic
function at one loop. Their existence is found to be
intimately related to the mechanism of anomaly cancellation.
The corrections found are exactly those that are expected by general
symmetry considerations \cite{N}. In (\ref{eq:f6f8}) we have given 
the result for the standard embedding. Coefficients might vary for 
more general cases, but the fact that they have opposite sign for 
the two separate groups is true in all known cases.

%
%
\subsection{Beyond one loop}

Not much can be said about the K\"ahler potential beyond one loop.
Superpotential and $f$--function, however, should not receive
further perturbative corrections. This implies that the
knowledge of $f$ at one loop represents the full perturbative
result. Combined with the fact that the coefficients are
fixed by anomaly considerations one would then expect that this
result for the $f$--function might be valid even beyond the
weakly coupled limit.

%
%
\section{Heterotic M--Theory}

Let us now review the calculation of 
\cite{NOY}. 
In the strongly coupled case we perform a compactification from $d=11$ 
to $d=4$ using the method of reduction and truncation as above. 
For the metric we write
\be
g^{(11)}_{MN} =
\left(
\ba{ccc}
c_4 e^{-\gamma} e^{-2\sigma} g_{\mu\nu} & & \\
 & e^\sigma g_{mn} & \\
 & & e^{2\gamma} e^{-2\sigma}
\ea
\right)
\label{eq:g11}
\ee
with 
$M,N = 1 \ldots 11$; $\mu,\nu = 1 \ldots 4$; $m,n = 5 \ldots 10$ 
and det($g_{mn}$)=1. 
This is the frame in which the 11--dimensional Einstein action 
gives the ordinary Einstein action after the reduction do $d=4$:
\be
-\frac{1}{2\kappa^2} \int d^{11}x \sqrt{g^{(11)}} R^{(11)} 
=
-\frac{c_4 \hat{V_7}}{2\kappa^2} \int d^{4}x \sqrt{g} R + \ldots
\ee
where $\hat{V_7}=\int d^7x$ is the coordinate volume of the 
compact 7--manifold and the scaling factor $c_4$ describes our 
freedom to choose the units in $d=4$. The most popular choice in the 
literature is $c_4=1$. This however corresponds to the unphysical 
situation in which the 4--dimensional Planck mass is determined 
by the choice of $\hat{V_7}$ which is just a convention. With 
$c_4=1$ one needs further rescaling of the 4--dimensional metric. 
We instead prefer the choice
\be
c_4 = V_7 / \hat{V_7}
\ee
where $V_7 = \int d^7x \sqrt{g^{(7)}}$ is the physical volume 
of the compact 7--manifold. This way we recover eq.\ (\ref{eq:GN}) 
in which the 4--dimensional Planck mass depends on the physical, 
and not coordinate, volume of the manifold on which we compactify. 
As a result, 
if we start from the product of the 4--dimensional Minkowski 
space and some 7--dimensional compact space  
(in the leading order of the expansion in $\kappa^{2/3}$) 
as a ground state in $d=11$ 
we obtain the Minkowski space with the standard normalization 
as the vacuum in $d=4$.

To find a more explicit formula for $c_4$ we have to discuss 
the fields $\sigma$ and $\gamma$ in some detail. In the 
leading approximation $\sigma$ is the 
overall modulus of the Calabi--Yau 6--manifold. We can divide 
it into a sum of the vacuum expectation value, 
$\left< \sigma \right>$, and the fluctuation $\tilde\sigma$.
In general both parts could depend on all 11 coordinates 
but in practice we have to impose some restrictions. 
The vacuum expectation value can not depend on $x^\mu$ 
if the 4--dimensional theory is to be Lorentz--invariant. 
In the fluctuations we drop the dependence on the compact 
coordinates corresponding to the higher Kaluza--Klein modes. 
Furthermore, we know that in the leading approximation 
$\left< \sigma \right>$ is just a constant, $\sigma_0$ , 
while corrections depending on the internal coordinates,
$\sigma_1$, are of the next order in $\kappa^{2/3}$. 
Thus, we obtain
\be
\sigma(x^\mu,x^m,x^{11})
=
\left<\sigma\right>(x^m,x^{11})
+ \tilde{\sigma}(x^\mu)
=
\sigma_0 
+ \sigma_1(x^m,x^{11}) 
+ \tilde{\sigma}(x^\mu)
\,.
\label{eq:sigma}
\ee
To make the above decomposition unique we define $\sigma_0$ 
by requiring that the integral of $\sigma_1$ over the internal 
space vanishes. The analogous decomposition can be also done 
for $\gamma$. With the above definitions the physical volume 
of the compact space is
\be
V_7 
= \int d^7x \left<e^{2\sigma}e^{\gamma}\right>
= e^{2\sigma_0} e^{\gamma_0} \hat{V_7}
\ee
up to corrections of order $\kappa^{4/3}$. Thus, the parameter  
$c_4$ can be written as
\be
c_4 = e^{2\sigma_0}e^{\gamma_0}
\,.
\ee

The choice of the coordinate volumes is just a convention. 
For example in the case of the Calabi--Yau 6--manifold only 
the product $e^{3\sigma}\hat{V_6}$ has physical meaning. 
For definiteness we will use the convention that the coordinate 
volumes are equal 1 in $M_{11}$ units. Thus, 
$\left<e^{3\sigma}\right>$ describes 
the Calabi--Yau volume in these units. 
Using eqs.\ (\ref{eq:VM11},\ref{eq:R11M11}) 
we obtain $e^{3\sigma_0} = V M_{11}^6 \approx (2.3)^6$, 
$e^{\gamma_0} e^{-\sigma_0} = R_{11} M_{11} \approx 9.2 a^2$. 
The parameter $c_4$ is equal to the square of the 4--dimensional 
Planck mass in these units and numerically $c_4 \approx (35a)^2$.

%
%
\subsection{Leading order}

At the classical level we compactify on 
$M^4 \times X^6 \times S^1/Z_2$. 
This means that the vacuum expectation values 
$\left<\sigma\right>$ and $\left<\gamma\right>$ 
are just constants and eq.\ (\ref{eq:sigma}) reduces to
\be
\sigma = \sigma_0 + \tilde\sigma(x^\mu)
,\qquad\qquad
\gamma = \gamma_0 + \tilde\gamma(x^\mu)
\,.
\ee
In such a situation $\sigma$ and $\gamma$ are 4--dimensional fields.
We introduce two other 4--dimensional fields by the relations
\bea
\frac{1}{4! c_4} e^{6\sigma} G_{11\lambda\mu\nu} &=& 
\epsilon_{\lambda\mu\nu\rho}\left(\partial^\rho D \right)
\,,
\\
C_{11 a {\bar b}} &=& C_{11} \delta_{a \bar b}
\eea
where $x^a$ ($x^{\bar b}$) is the holomorphic (antiholomorphic) coordinate 
of the Calabi--Yau manifold. 
Now we can define the dilaton and the modulus fields by
\bea
\sS &=& \frac{1}{\left( 4\pi \right)^{2/3}} 
\left( e^{3\sigma} + i 24 \sqrt{2} D \right)
\,,
\label{eq:sS0}
\\
\sT &=& \frac{1}{\left( 4\pi \right)^{2/3}}
\left( e^{\gamma} + i 6 \sqrt{2} C_{11} + C^*_i C_i \right)
\label{eq:sT0}
\eea
where the observable sector matter fields $C_i$ originate from the
gauge fields $A_M$ on the 10--dimensional observable wall 
(and $M$ is an index in
the compactified six dimensions).
The K\"ahler potential takes its standard form as in eq.\ (\ref{eq:G}) 
\be
K = - \log (\sS + \sS^*) - 3 \log (\sT+\sT^* -2 C^*_i C_i)
\label{eq:sK}
\,.
\ee
The imaginary part of $\sS$ (Im$\sS$) corresponds to the model 
independent axion, and with the above normalization the gauge kinetic 
function is $f = \sS$. We have also
\be
W(C) = d_{ijk} C_iC_jC_k   
\ee
This is very similar to the weakly coupled case.

%
%
\subsection{Next to leading order}

Before drawing any conclusion from the formulae obtained above we 
have to discuss a possible obstruction at the next to leading order. 
For the 3--index tensor field $H$ in $d=10$ supergravity to be well 
defined one has to satisfy $dH = \tr F_1^2 + \tr F_2^2 - \tr R^2 = 0$ 
cohomologically. In the simplest case of the standard embedding one
assumes $\tr F_1^2 = \tr R^2$ locally and the gauge group is broken to 
$E_6 \times E_8$. Since in the M--theory case the two different gauge 
groups live on the two different boundaries (walls) of space--time 
such a cancellation point by point is no longer possible 
\cite{W}.
We expect nontrivial vacuum expectation values (vevs) of
\be
(dG) \propto \sum_i \delta(x^{11} - x^{11}_i) 
\left( \tr F_i^2 - {1\over 2} \tr R^2 \right)
\ee
at least on one boundary ($x^{11}_i$ is the position of $i$--th 
boundary). In the case of the standard embedding we would have 
$\tr F_1^2 - {1\over 2} \tr R^2 = {1\over 2} \tr R^2$ on one and 
$\tr F^2_2 - {1\over 2} \tr R^2 = - {1\over 2} \tr R^2$ on the other
boundary. This might pose a severe problem since a nontrivial vev  of 
$G$ might be in conflict with supersymmetry ($G_{11ABC}=H_{ABC}$). 
The supersymmetry transformation law in $d=11$ reads  
\be
\delta \psi_M
=
D_M\eta + \frac{\sqrt{2}}{288} G_{IJKL} 
          \left( \Gamma_M^{IJKL} - 8 \delta_M^I \Gamma^{JKL} 
            \right) \eta
+ \ldots
\label{eq:dpsiM}
\ee
Supersymmetry will be broken unless e.g.\ the derivative term 
$D_M\eta$ compensates the nontrivial vev of $G$. Witten has shown 
\cite{W} 
that such a cancellation can occur and constructed the solution in 
the linearized approximation 
(linear in the expansion parameter $\kappa^{2/3}$)\footnote
{For a discussion beyond this approximation in the weakly coupled 
case see ref.\ 
\cite{NS}.
}.
This solution requires some modification of the metric on $M^{11}$:
\be
g^{(11)}_{MN} =
\left(
\ba{ccc}
(1+b) \eta_{\mu\nu} & & \\
 & (g_{ij}+h_{ij}) & \\
 & & (1+\gamma') 
\ea
\right)
\,.
\label{eq:gW}
\ee
$M^{11}$ is no longer a direct product $M^4 \times X^6 \times S^1/Z_2$ 
because $b$, $h_{ij}$ and $\gamma'$ depend
now on the compactified coordinates.
The volume of $X^6$ depends on $x^{11}$ 
\cite{W}:
\be
\frac{\partial}{\partial x^{11}} V 
=
-\frac{\sqrt{2}}{8}
{\int d^6x \sqrt{g} \omega^{AB}\omega^{CD}G_{ABCD}}
\label{eq:d11V_W}
\ee
where the integral is over the Calabi--Yau manifold $X^6$ and $\omega$
is the corresponding K\"ahler form.
The parameter $(1+b)$ is the scale factor of
the Minkowski 4--manifold and depends on $x^{11}$ in the following way
\be
\frac{\partial}{\partial x^{11}} b =
\frac{1}{2}  \frac{\partial}{\partial x^{11}} \log v_4 =
\frac{\sqrt{2}}{24}
\omega^{AB}\omega^{CD}G_{ABCD}
\label{eq:d11b}
\ee
where $v_4$ is the physical volume for some fixed coordinate volume
in $M^4$.
In our simple reduction and truncation method
with the metric $g^{(11)}_{MN}$ given by eq.\ (\ref{eq:g11}) 
we can reproduce the $x^{11}$ dependence of $V$ and $v_4$.
The volume of $X^6$ is determined by $\sigma$:
\be
\frac{\partial}{\partial x^{11}} \log V 
= \frac{\partial}{\partial x^{11}} \left(3 \left<\sigma\right>\right)
= 3 \frac{\partial}{\partial x^{11}} \sigma
\label{eq:d11V}
\ee
while the scale factor of $M^4$ can be similarly
expressed in terms of $\sigma$ and $\gamma$ fields:
\be
\frac{\partial}{\partial x^{11}} \log v_4 
= - \frac{\partial}{\partial x^{11}} 
\left(2 \left<\gamma\right> + 4 \left<\sigma\right>\right)
= - \frac{\partial}{\partial x^{11}} (2 \gamma + 4 \sigma)
\label{eq:d11v4}
\,.
\ee
Substituting $\left<\sigma\right>$ with $\sigma$ in the above 
two equations is allowed because, due to our decomposition 
(\ref{eq:sigma}), only the vev of $\sigma$ depends on the 
internal coordinates (the same is true for $\gamma$).
The scale factor $b$ calculated in ref.\ \cite{W}
depends also on the Calabi--Yau coordinates.
Such a dependence can not be reproduced in our simple
reduction and truncation compactification so we have to average
eq.\ (\ref{eq:d11b}) over $X^6$.
Using equations (\ref{eq:d11V_W}--\ref{eq:d11v4})
after such an averaging we obtain
(to leading order in the expansion parameter $\kappa^{2/3}$) 
\cite{NOY}
\be
\frac{\partial\gamma}{\partial x^{11}} 
=
-\frac{\partial\sigma}{\partial x^{11}} 
=
\frac{\sqrt{2}}{24}
\frac
{\int d^6x \sqrt{g} \omega^{AB}\omega^{CD}G_{ABCD}}
{\int d^6x \sqrt{g}}
\,.
\label{eq:sigmagamma1}
\ee
Substituting the vacuum expectation value of $G$ found in 
\cite{W}
we can rewrite it in the form
\be
\frac{\partial\gamma}{\partial x^{11}} 
=
-\frac{\partial\sigma}{\partial x^{11}} 
=
\frac{2}{3} \alpha \kappa^{2/3} V^{-2/3}
\label{eq:sigmagamma2}
\ee
where 
\be
\alpha = \frac{\pi c}{2(4\pi)^{2/3}}
\label{eq:alpha}
\ee
and c is a constant of order unity given for the standard embedding 
of the spin connection by
\be
c =  V^{-1/3}
\left| \int \frac{\omega \wedge \tr (R \wedge R)}{8 \pi^2} \right|
\,.
\label{eq:c}
\ee
Our calculations, as those of Witten, are valid only in the 
leading nontrivial order in the $\kappa^{2/3}$ expansion. 
The expression (\ref{eq:sigmagamma2}) for the derivatives 
of $\sigma$ and $\gamma$ have explicit factor $\kappa^{2/3}$. 
This means that we should take the lowest order value for the 
Calabi--Yau volume in that expression. An analogous procedure 
has been used in obtaining all formulae presented in this paper. 
We always expand in $\kappa^{2/3}$ and drop all terms which are 
of higher order than our approximation. Taking the above into 
account and using our units in which $M_{11}=1$ we can rewrite 
eq.\ (\ref{eq:sigmagamma2}) in the simple form:
\be
\frac{\partial\gamma}{\partial x^{11}} 
=
-\frac{\partial\sigma}{\partial x^{11}} 
=
\frac{2}{3} \alpha e^{-2\sigma_0}
\,.
\label{eq:sigmagamma3}
\ee

Eqs.\ (\ref{eq:sigmagamma1}--\ref{eq:sigmagamma3}) represent one 
of the main results of ref.\ \cite{NOY}. As we will see in the
following, this result contains all the information to deduce
the effective action, i.e. K\"ahler potential,
superpotential and gauge kinetic function of the 4--dimensional 
effective supergravity theory. 

As we showed in \cite{NOY},
it is the above dependence of $\sigma$ and $\gamma$ on $x^{11}$
that leads to these  consequences. 
One has to be careful in defining the fields in $d=4$. It is obvious,
that the 4--dimensional fields $\sS$ and $\sT$ can not be any 
longer defined by eqs.\ (\ref{eq:sS0}, \ref{eq:sT0}) 
because now $\sigma$ and $\gamma$ are 5--dimensional fields. 
We have to integrate out the dependence on the 11th coordinate. 
In the present approximation, this procedure is quite simple:
we have to replace $\sigma$ and $\gamma$ in the 
definitions of $\sS$ and $\sT$ with their averages over 
the $S^1/Z_2$ interval \cite{NOY}. 
With the linear dependence of $\sigma$ and $\gamma$ on $x^{11}$ 
their average values coincide with the values taken at the 
middle of the $S^1/Z_2$ interval
\be
\bar \sigma 
= \sigma \left( \frac{\pi\rho}{2} \right)
= \sigma_0 + \tilde{\sigma}(x^\mu)
\,,
\ee
\be
\bar \gamma 
= \gamma \left( \frac{\pi\rho}{2} \right)
= \gamma_0 + \tilde{\gamma}(x^\mu)
\,.
\ee

When we reduce the boundary part of the Lagrangian of M--theory 
to 4 dimensions we find exponents of $\sigma$ and $\gamma$ 
fields evaluated at the boundaries. Using eqs.\ (\ref{eq:sigma}) 
and (\ref{eq:sigmagamma3}) we get
\bea
e^{-\gamma} \big|_{M^{10}_i}
&=&
e^{-\gamma_0} \pm \frac{1}{3} \alpha e^{-3\sigma_0}
\,,
\label{eq:exp-gamma}
\\
e^{3\sigma} \big|_{M^{10}_i}
&=&
e^{3\sigma_0} \pm \alpha e^{\gamma_0}
\,.
\label{eq:exp3sigma}
\eea
The above formulae have very important consequences for 
the definitions of the K\"ahler potential and the 
gauge kinetic functions. For example, the coefficient in front of the 
$D_\mu C^*_i D^\mu C_i$ kinetic term is proportional to $e^{-\gamma}$ 
evaluated at the $E_6$ wall where the matter fields propagate. 
At the lowest order this was just $e^{-\gamma_0}$ or 
$\left<\sT\right>^{-1}$ up to some numerical 
factor. From eq.\ (\ref{eq:exp-gamma}) we see that 
at the next to leading order also $\left<\sS\right>^{-1}$ 
is involved with relative coefficient $\alpha/3$. 
Taking such corrections into account we find that 
at this order the K\"ahler potential is given 
by\footnote
{
Our solution is valid only for terms at most linear in $\alpha$. 
Keeping this in mind we could write the K\"ahler potential 
also in the form 
$
K= 
-\log(\sS + \sS^*- 2 \alpha C^*_i C_i)
- 3 \log (\sT+ \sT^* - 2 C^*_i C_i)
$.
} 
\be
K 
= 
- \log (\sS + \sS^*) 
+ \frac{2 \alpha C^*_i C_i}{\sS + \sS^*}
- 3 \log (\sT+ \sT^* - 2 C^*_i C_i)
\ee
with $\sS$ and $\sT$ now defined by
\bea
\sS &=& \frac{1}{\left( 4\pi \right)^{2/3}}
\left( e^{3\bar\sigma} + i 24 \sqrt{2} \bar D + \alpha C^*_i C_i \right)
\,,
\\
\sT &=& \frac{1}{\left( 4\pi \right)^{2/3}}
\left( e^{\bar\gamma} + i 6 \sqrt{2} \bar C_{11} + C^*_i C_i \right)
\label{eq:sT}
\eea
where bars denote averaging over the 11th dimension. 
It might be of some interest to note that the combination
$\left<\sS\right>\left<\sT\right>^3$ is independent of $x^{11}$ 
even before this averaging procedure took place.

Equipped with this definition the calculation of the gauge kinetic
function(s) from eqs.\ (\ref{eq:sigmagamma3}, \ref{eq:exp3sigma})
becomes a trivial exercise \cite{NOY}. In the five--dimensional
theory $f$ depends on the 11--dimensional coordinate as well, thus the 
gauge kinetic function takes different values at the two walls. 
The averaging procedure allows us to deduce these functions directly. 
For the simple case at hand (the so--called standard embedding)
eq.\ (\ref{eq:exp3sigma}) gives \cite{NOY}
\be
f_6 = \sS +\alpha \sT\,;
\qquad\qquad
f_8 = \sS - \alpha \sT\,.
\label{eq:f6f8alpha}
\ee
It is a special property of the standard embedding that the 
coefficients are equal and opposite. The coefficients vary for 
more general cases. This completes the discussion of the $d=4$ 
effective action in next to leading order, noting that the 
superpotential does not receive corrections at this level.

The nontrivial dependence of $\sigma$ and $\gamma$ on $x^{11}$
can also enter definitions and/or interactions of other
4--dimensional fields. Let us next consider the gravitino. 
After all we have to show that this field is massless to give 
the final proof that the given solution respects supersymmetry.
Its 11--dimensional kinetic term
\be
- \frac{1}{2} \sqrt{g} \psib_I \Gamma^{IJK} D_J \psi_K
\label{eq:grav_kin}
\ee
remains diagonal after compactification to $d=4$ if we
define the 4--dimensional gravitino, $\psi^{(4)}_\mu$,
and dilatino ,$\psi^{(4)}_{11}$, fields by the relations
\bea
\psi_\mu
&=&
e^{-(\sigma-\sigma_0)/2}e^{-(\gamma-\gamma_0)/4}
\left(\psi^{(4)}_\mu + \frac{1}{\sqrt{6}}\Gamma_\mu\psi^{(4)}_{11}\right)
\,,
\label{eq:grav_def1}
\\
\psi_{11}
&=&
-\frac{2}{\sqrt{6}}e^{(\sigma-\sigma_0)/2} e^{(\gamma-\gamma_0)/4}
\Gamma^{11} \psi^{(4)}_{11}
\,.
\label{eq:grav_def2}
\eea
The $d=11$ kinetic term (\ref{eq:grav_kin}) gives after
the compactification also a mass term for the $d=4$ gravitino
of the form
\be
\frac{3}{8} e^{\sigma_0} e^{-\gamma_0}
\frac{\partial\gamma}{\partial x^{11}}
=
\frac{\sqrt{2}}{64} e^{\sigma_0} e^{-\gamma_0}
\frac
{\int d^6x \sqrt{g} \omega^{AB}\omega^{CD}G_{ABCD}}
{\int d^6x \sqrt{g}}
=
\frac{1}{4} \alpha e^{-\sigma_0} e^{-\gamma_0}
\,.
\label{eq:grav_mass}
\ee
The sources of such a term are nonzero values of the spin 
connection components $\omega_\mu^{\alpha 11}$ and 
$\omega_m^{a 11}$ resulting from the $x^{11}$ dependence of 
the metric. It is a constant mass term from the 4--dimensional 
point of view. This, however, does not mean that the gravitino 
mass is nonzero. There is another contribution from the 
11--dimensional term
\be
- \frac{\sqrt{2}}{384} \sqrt{g}
\psib_I \Gamma^{IJKLMN} \psi_N
\left( G_{JKLM} + {\hat G}_{JKLM} \right)
\,.
\label{eq:GG}
\ee
After redefining fields according to
(\ref{eq:grav_def1},\ref{eq:grav_def2})
and averaging the nontrivial vacuum expectation value of $G$
over $X^6$ we get from eq.\ (\ref{eq:GG})
a mass term which exactly cancels the previous contribution
(\ref{eq:grav_mass}).
The gravitino is massless -- the result which we expect
in a model with unbroken supersymmetry and vanishing
cosmological constant.
Thus, we find that our simple reduction and truncation method
(including the correct $x^{11}$ dependence in next to leading order)
reproduces the main features of the model.

%
%
\subsection{Critical radius}

The factor $\left<\exp(3\sigma)\right>$ represents the  volume of the
six--dimensional compact space in units of $M_{11}^{-6}$.
The $x^{11}$ dependence of $\sigma$
then leads to the geometrical picture that the volume of this space
varies with $x^{11}$ and differs at the two boundaries:
\be
V_{E_8}
=
V_{E_6} - 2 \pi^2 \rho \left(\frac{\kappa}{4 \pi}\right)^{2/3}
\left|
\int \omega \wedge
     \frac{\tr (F \wedge F) - \frac{1}{2} \tr (R \wedge R)}{8 \pi^2}
\right|
\ee
where the integral is over $X^6$ at the $E_6$ boundary.
In the given approximation, this variation is linear,
and for growing $\rho$ the volume on the $E_8$ side becomes
smaller and smaller. At a critical value of $\rho$
the volume will thus vanish and this will provide
us with an upper limit on $\rho$:
\be
\rho
<
\rho_{crit}
=
\frac{(4 \pi)^{2/3}}{c\pi^2} M_{11}^{3} V_{E_6}^{2/3}
\label{eq:rho}
\ee
where $c$ was defined in eq.\ (\ref{eq:c}). 
To estimate the numerical value of $\rho_{crit}$ we first recall
that from eq.\ (\ref{eq:GN}) we obtained\footnote
{With $V$ depending on $x^{11}$ we have to specify which values
should be used in 
eqs.\ (\ref{eq:GN},\ref{eq:alphaGUT},\ref{eq:VMGUT}). 
The appropriate choice in the expression for $G_N$ is the average 
value of $V$ while in the expressions for $\alpha_{GUT}$ 
and for the $V$--$M_{GUT}$ relation we have to use $V$ 
evaluated at the $E_6$ wall.}
\be
M_{11} V_{E_6}^{1/6}
=
\left(\alpha_{GUT} (4 \pi)^{-2/3} \right)^{-1/6} \approx 2.3
\,.
\ee
Thus, we get
\be
\rho^{-1}
>
\rho_{crit}^{-1}
\approx
0.16 c V_{E_6}^{-1/6}
\,.
\label{eq:rho_crit}
\ee
The numerical value of $V$ at the $E_6$ boundary
depends on what we identify with the unification scale $M_{GUT}$ 
via eq.\ (\ref{eq:VMGUT}):
\be
V_{E_6}^{-1/6}
=
aM_{GUT}^{-1}
\ee
with $a$ somewhere between 1 and about 2. 
Thus, the bound (\ref{eq:rho_crit}) can be written in the form
\be
R_{11}^{-1}
>
0.05 \frac{c}{a} M_{GUT}
\,.
\label{eq:R11_crit}
\ee

For the phenomenological applications we have to check
whether our preferred choice of
$6.2 \cdot 10^{14}\,\, \GeV < R_{11}^{-1} < 7.4 \cdot 10^{15}$ GeV 
that fits the correct value of the $d=4$ Planck mass
satisfies the bound (\ref{eq:R11_crit}).
In a rather extreme case of $c=1$ and $a=2.3$ 
we find that the upper bound on $R_{11}^{-1}$ is of the order of
$6.5 \cdot 10^{14}$ GeV. Even for $c=1$ this bound goes up to about
$1.5 \cdot 10^{15}$ GeV if we identify $V^{-1/6}$ with $M_{GUT}$.
Although some coefficients are model dependent we find in general 
that the bound can be satisfied, but that $R_{11}$ is
quite close to its critical value. Values of $R_{11}^{-1}$ about 
$10^{12}$ GeV as necessary in \cite{AQ} seem to be beyond the 
critical value, even with the modifications discussed at the end 
of Chapter 2. In any case, models where supersymmetry is broken 
by a Scherk--Schwarz mechanism seem to require the absence of the 
next to leading order corrections in (\ref{eq:f6f8alpha}), 
i.e.\ $\alpha=0$. It remains to be seen whether such a possibility 
can be realized.

%
%
\subsection{Connection to the weakly coupled case}

Inspection of (\ref{eq:f6f8}) and (\ref{eq:f6f8alpha}) reveals a close
connection between the strongly and weakly coupled case \cite{BD,NS}.
The variation of the Calabi--Yau manifold volume as discussed above
is the analogue of the one loop correction of the gauge kinetic 
function (\ref{eq:f6f8}) in the weakly coupled case and has the same 
origin, namely a Green--Schwarz anomaly cancellation counterterm. 
In fact, also in the strongly coupled case this leads to a correction
for the gauge coupling constants at the $E_6$ and $E_8$ side. 
As seen, gauge couplings are no longer given by the (averaged) 
$\sS$--field, but by that combination of the (averaged) $\sS$ and 
$\sT$ fields which corresponds to the $\sS$--field before averaging 
at the given boundary leading to 
\be
f_{6,8} = \sS \pm \alpha \sT
\ee
at the $E_6$ ($E_8$) side respectively.
The critical value of $R_{11}$ will correspond to infinitely strong
coupling at the $E_8$ side $\sS - \alpha \sT = 0$.
Since we are here close to criticality a correct phenomenological  
fit of $\alpha_{\rm GUT} = 1/25$ should include this correction
$\alpha_{\rm GUT}^{-1} = \sS + \alpha \sT$ where $\sS$ and
$\alpha \sT$ give comparable contributions. This is a difference to 
the weakly coupled case, where in $f= \wS + \epsilon \wT$ the latter
contribution was small compared to $\wS$. This stable result for the 
corrections to $f$ when going from weak coupling to strong coupling 
is only possible because of the rather special properties of $f$. 
$f$ does not receive further perturbative corrections beyond one loop
\cite{SV,N}, and the one loop corrections are determined by
the anomaly considerations. The formal expressions for the
corrections are identical, the difference being only that in the
strongly coupled case these corrections are as important as
the classical value.

%
%
\section{Supersymmetry breaking at the hidden wall}

We shall now discuss the question of supersymmetry breakdown 
within this framework. We consider the breakdown of
supersymmetry in a hidden sector, transmitted to the observable
sector via gravitational interactions. Such a scenario was suggested
in \cite{HPN2} after having observed that gaugino condensation
can break supersymmetry in $d=4$ supergravity models. A nontrivial
gauge kinetic function $f$ seems to be necessary for such
a mechanism to work \cite{FGN}. In the heterotic string both
ingredients, a hidden sector $E_8$ and a nontrivial $f$, were
present in a natural way and a coherent picture of supersymmetry
breakdown via gaugino condensation emerged \cite{DIN,DRSW,DIN2}.
In the strongly coupled case, such a mechanism can be realized as 
well \cite{H,NOY}. In fact the notion of the hidden sector acquires
a geometrical interpretation: the gaugino condensate forms at one 
boundary (the hidden wall) of spacetime. We shall now discuss
this mechanism in detail. First we review the situation
in the weakly coupled case. 
  Our aim then is to compare the
strong coupling regime with the weak coupling regime and clarify
similarities as well as differences. 

%
%
\subsection{Weak coupling case}

Let us first discuss supersymmetry breaking in the weakly coupled
case using the action of $d=10$ supergravity. 
Supersymmetry transformation laws for the $d=10$ gravitino
fields $\psi_M$ and the dilatino field $\lambda$ are
written\footnote{Here we use the conventions of \cite{CM}, 
where the Lagrangian is given in the Einstein frame. 
To recover the effective action (\ref{eq:10d+H}) in the string frame, 
one has to make a proper Weyl transformation and identify 
$\varphi=\exp{(\phi/3)}$.}
\begin{eqnarray}
  \delta \lambda &=& \frac{1}{8} \varphi^{-3/4} \Gamma^{MNP} H_{MNP}
    +\frac{\sqrt{2}}{384} \Gamma^{MNP} \bar \chi^a \Gamma_{MNP} \chi^a
    +\ldots  , \nonumber \\
  \delta \psi_M &=& \frac{\sqrt{2}}{32} \varphi^{-3/4} 
             (\Gamma_{M}^{NPQ} - 9 \delta_M^N \Gamma^{PQ}) H_{NPQ} 
\nonumber \\
  & &    +\frac{1}{256} (\Gamma_{MNPQ}-5 g_{MN}\Gamma_{PQ}) 
       \bar \chi^a \Gamma^{NPQ} \chi^a +\ldots,
\end{eqnarray}
implying that a condensate of gauginos $\bar \chi \chi$ and/or
non--vanishing vevs of the $H$ fields may break supersymmetry. Here we
assume the appearance of the gaugino condensate in the hidden sector
\begin{equation}
   \langle \bar \chi^a \Gamma_{mnp} \chi^a \rangle =\Lambda ^3  
                                                      \epsilon_{mnp},
\end{equation}
with $\Lambda$ being the gaugino condensation scale and
$\epsilon_{mnp}$ the covariantly constant holomorphic three--form.  
The perfect square structure seen in the Lagrangian \cite{DRSW}
\begin{equation}
  -\frac{3}{4} \varphi^{-3/2} ( H_{MNP} -\sqrt{2}\varphi^{3/4} 
         \bar \chi^a \Gamma_{MNP} \chi^a )^2
\label{eq:perfect-square-weak}
\end{equation}
will be a very important ingredient to discuss the quantitative
properties of the mechanism. When reducing to the $d=4$ effective
action we will find a cancellation of the vevs of the $H$ field
and the gaugino condensate at the minimum of the potential
such that the term in eq.~(\ref{eq:perfect-square-weak}) vanishes. 
Before we look at this in detail, let us first comment on such
a possible vev of $H$ and a possible quantization condition
of the antisymmetric tensor. In \cite{RW} it was shown, that
an antisymmetric tensor field $H=dB$ has a quantized vacuum
expectation value. In many subsequent papers this has been
incorrectly taken as an argument for the quantization of
the vev of $H=dB+\omega^{YM}-\omega^{L}$ as given in
eq.~(\ref{eq:Hfield}). The correct way to interpret this
situation is to have a cancellation of the gaugino condensate
with the vev of a Chern--Simons term \cite{DIN2}, for which
such a quantization condition does not hold. After all the
Chern--Simons term $\omega^{YM}$ contains the superpotential of
the $d=4$ effective theory \cite{DIN}.
This cancellation leads to a certain combination of
$\psi_M$ and $\lambda$ as the candidate goldstino that will
provide the longitudinal component of the gravitino. While in
$d=10$ this looks rather complicated, it simplifies 
tremendously once one reduces to $d=4$. Qualitatively the
scalar potential takes the following form 
at the classical level (for the detailed
factors see \cite{HPN3}):
\be
V={1\over{ST^3}}\left[ \mid W - 2(ST)^{3/2}(\bar\chi\chi)\mid^2
+ {T\over 3}\mid{\partial W\over{\partial C}}\mid^2\right].
\label{eq:potential}   
\ee
We observe the important fact that the potential is positive and
vanishes at the minimum. Thus we have broken supersymmetry with a 
vanishing cosmological constant at the classical level.
The first term in the brackets of eq.~(\ref{eq:potential})
corresponds to the contribution from eq.~(\ref{eq:perfect-square-weak})
once reduced to $d=4$ and vanishes at the minimum. In the $d=4$
theory it represents the auxiliary component $F_S$ of the dilaton
superfield $S$. Thus we have $F_S=0$ and supersymmetry is
broken by a nonvanishing vev of $F_T$ \cite{DIN2}. The goldstino is
then the fermion in the $T$--multiplet and we are dealing with a 
situation that has later been named moduli--dominated supersymmetry
breakdown. This fact has its origin in the special properties
of the $d=10$ action (the term in eq.~(\ref{eq:perfect-square-weak}))
and seems to be of rather general validity. The statement
$F_S=0$ is, of course, strictly valid only in the classical theory.
The corrections discussed in section 3, eq.~(\ref{eq:f6f8}) will
slightly change these results as we shall discuss in section 6.

Having minimized the potential and identified the goldstino we can
now compute the gravitino mass according to the standard procedure.
The result has a direct physical meaning because we are dealing
with a theory with vanishing vacuum energy. We obtain
\begin{equation}
   m_{3/2} \sim \frac{F_T}{M_{Planck}} 
\sim \frac{\Lambda^3}{M_{Planck}^2}.
\label{eq:gravmass}
\end{equation}
A value of $\Lambda \sim 10^{13}$ GeV will thus lead to a 
gravitino mass in the TeV region. So far is our review of the
mechanism of gaugino condensation in the weakly coupled theory.

%
%
\subsection{Strongly coupled case}

Next we turn to supersymmetry breaking in the strongly coupled case
($d=11$ M--theory picture) and start with the $d=11$ action.  
Supersymmetry transformation laws for the
gravitino fields in this case are given by
\begin{eqnarray}
  \delta \psi_A &=& D_A\eta + 
      \frac{\sqrt{2}}{288} G_{IJKL} \left(
    \Gamma_A^{IJKL} - 8 \delta_A^I \Gamma^{JKL} \right) \eta \nonumber
  \\ 
    &&- \frac{1}{1152\pi} \left( \frac{\kappa}{4\pi} \right)^{2/3}
  \delta(x^{11}) \left( \bar \chi^a \Gamma_{BCD} \chi^a \right) \left
    ( \Gamma_A^{BCD} - 6 \delta_A^B \Gamma^{CD} \right) \eta + \ldots
  \label{eq:susytr-strong-A} \\
  \delta \psi_{11} &=& D_{11} \eta + \frac{\sqrt{2}}{288} G_{IJKL}
  \left( \Gamma_{11}^{IJKL} - 8 \delta_{11}^I \Gamma^{JKL} \right)
  \eta \nonumber \\ &&+ \frac{1}{1152\pi} \left( \frac{\kappa}{4\pi}
  \right)^{2/3} \delta(x^{11}) \left( \bar \chi^a \Gamma_{ABC} \chi^a
  \right) \Gamma^{ABC} \eta + \ldots \label{eq:susytr-strong-B}
\end{eqnarray}
where gaugino bilinears appear in the right hand side of both
expressions. Again we consider gaugino condensation  at the 
hidden $E_8$ boundary 
\begin{equation}
\langle \bar{\chi}^a \Gamma_{ijk} \chi^a \rangle = g_8^2 \Lambda^3
\epsilon_{ijk}.
\end{equation}
The $E_8$ gauge coupling constant appears in this equation because 
the straightforward reduction and truncation leaves a non--canonical
normalization for the gaugino kinetic term. 
An important property of the  weakly coupled case (d=10
Lagrangian) was the fact that  the gaugino condensate and 
the three--index tensor field
$H$ contributed to the scalar potential in a full square.  
Ho\v{r}ava made the important observation that a
similar structure appears in the M--theory Lagrangian as well \cite{H}:
\begin{equation}
- \frac{1}{12\kappa^2} \int_{M^{11}} d^{11}x \sqrt{g}
  \left(G_{ABC11} 
     - \frac{\sqrt{2}}{32\pi} \left( \frac{\kappa}{4\pi} \right)^{2/3}
               \delta(x^{11}) \bar{\chi}^a \Gamma_{ABC} \chi^a
  \right)^2  \label{eq:perfect-square-strong}
\end{equation}
with the obvious relation between $H$ and $G$. Let us now have a closer
look at the form of $G$. At the next to leading order we have
\begin{eqnarray}
     G_{11ABC}&=&(\partial_{11} C_{ABC} +\mbox{permutations}) 
\nonumber \\
             & &+\frac{1}{4 \pi \sqrt{2}} 
              \left( \frac{\kappa}{4 \pi} \right)^{2/3}
             \sum_{i} \delta(x^{11}-x_i^{11})
          ( \omega^{YM}_{ABC}-\frac{1}{2} \omega^L_{ABC} ).
\end{eqnarray}
Observe, that in the bulk we have $G=dC$ with the Chern--Simons
contributions confined to the boundaries.  Formula 
(\ref{eq:perfect-square-strong}) suggests a cancellation between 
the gaugino condensate and the $G$--field in a way very similar to 
the weakly coupled case, but the nature of the cancellation of the 
terms becomes much more transparent now. In the former case we had 
to argue via the quantization condition for $dB$ that the gaugino 
condensate is cancelled by one of the Chern--Simons terms. Here this 
becomes obvious. The condensate is located at the wall as are the 
Chern--Simons terms, so this cancellation has to happen 
{\bf locally at the wall}. A quantization condition for $dC$ 
(the generalization of the quantization condition for $dB$) 
has been discussed in ref.\ \cite{WQ}.

So this cancellation is very similar to the one in the weakly 
coupled case. At the minimum of the potential we obtain 
\begin{equation}
G_{ABC11} 
     = \frac{\sqrt{2}}{32\pi} \left( \frac{\kappa}{4\pi} \right)^{2/3}
               \delta(x^{11}) \bar{\chi}^a \Gamma_{ABC} \chi^a
\label{eq:vev-G}
\end{equation}
at the hidden wall. Substituting this into 
eqs.\ (\ref{eq:susytr-strong-A}) and (\ref{eq:susytr-strong-B}) 
and using Witten's solution \cite{W} (as discussed in section 4.2) 
we obtain 
\be
\delta \psi_{11}
=  
\frac{1}{384\pi} \left( \frac{\kappa}{4\pi}\right)^{2/3} \delta(x^{11}) 
\left( \bar \chi^a \Gamma_{ABC} \chi^a \right) \Gamma^{ABC} \eta 
+ \ldots  
\,. 
\label{eq:susy-tr}
\ee
This nonzero expectation value of $\delta \psi_{11}$ shows that 
supersymmetry is spontaneously broken.
Because of the cancellation in eq.~(\ref{eq:perfect-square-strong}), 
the cosmological constant vanishes at leading order.  
Recalling supersymmetry transformation law for the elfbein
\begin{equation}
    \delta e^m_I=\frac{1}{2}\bar \eta \Gamma^m \psi_I,
\end{equation}
one finds that the superpartner of the $\sT$ field plays the role 
of the goldstino. Again we have a situation where $F_{\sS}=0$ 
(due to the cancellation in (\ref{eq:perfect-square-strong})) with
nonvanishing $F_{\sT}$. But here we find the novel and interesting
situation that $F_{\sT}$ differs from zero only at the hidden wall,
although the field itself is a bulk field\footnote{In general
it would be interesting
to consider also situations where the goldstino is not a bulk
but a wall field.}.
At that wall our discussion is  completely 4--dimensional although
we are still dealing effectively with a $d=5$ theory. To reach
the effective theory in $d=4$ we have to integrate out the 
dependence of the $x^{11}$ coordinate. As in the previous section
this can be performed by the averaging procedure explained there.
With the gaugino condensation scale $\Lambda$ sufficiently small 
compared to the
compactification scale $M_{GUT}$, the low--energy effective theory is
well described by four dimensional $N=1$ supergravity in which
supersymmetry is spontaneously broken.  In this case, the modes which
remain at low energies will be well approximated by constant modes
along the $x^{11}$ direction.  This observation justifies our
averaging procedure to obtain four dimensional quantities. 
Averaging $\delta \psi_{11}$ over $x^{11}$, we thus obtain the vev
of the auxiliary field $F_{\sT}$
\begin{equation}
    F_{\sT}=\frac{1}{2} {\sT} 
         \frac{\int dx^{11} \sqrt{g_{1\!1 1\!1}} \delta \psi_{11}}
              { \int dx^{11} \sqrt{g_{1\!1 1\!1}}}.
\end{equation}
Note that this procedure allows for a nonlocal cancellation of the
vev of the auxiliary field in $d=4$. A condensate with equal
size and opposite sign at the observable wall could cancel the
effect and restore supersymmetry.
Using $\int dx^{11} \sqrt{g_{1\!1 1\!1}} \delta (x^{11})=1$, the 
auxiliary field is
found to be
\begin{equation}
  F_{\sT} ={\sT} \frac{1}{32 \pi (4 \pi)^{2/3}} 
           \frac{ g_8^2 \Lambda^3}{R_{11} M_{11}^3 }
\end{equation}
Similarly one can easily show that $F_{\sS}$ as well as the
vacuum energy vanish. This allows us then to unambiguously determine
the gravitino mass, which is related to the auxiliary field in
the following way:
\begin{equation}
  m_{3/2}=\frac{F_{\sT}}{{\sT}+{\sT}^*} 
 =\frac{1}{64\pi (4 \pi)^{2/3}} \frac{g_8^2 \Lambda^3}{R_{11} M_{11}^3}
 =\frac{\pi}{2} \frac{\Lambda^3}{M_{Planck}^2}.
\label{eq:gravitino-mass}
\end{equation}
As a nontrivial check one may calculate the gravitino mass 
in a different way. A term in the 
Lagrangian
\begin{equation}
  -\frac{\sqrt{2}}{192 \kappa^2}
   \int dx^{11} \sqrt{g} \bar \psi_I \Gamma^{IJKLMN} \psi_N 
                G_{JKLM},
\end{equation}
becomes the gravitino mass term when compactified to four dimensions.
Using the vevs of the $G_{IJK11}$ given by eq. (\ref{eq:vev-G}), one
can obtain the same result as eq.~(\ref{eq:gravitino-mass}). This is a
consistency check of our approach and the fact that the vacuum
energy vanishes in the given approximation.

It follows from eq.~(\ref{eq:gravitino-mass}), that the gravitino mass
tends to zero when the radius of the eleventh dimension goes to 
infinity. When the four--dimensional Planck scale is fixed to be 
the measured value, however, the gravitino mass in the strongly 
coupled case is expressed in a standard manner, similar to the weakly 
coupled case as can be seen by inspecting (\ref{eq:gravitino-mass})
and (\ref{eq:gravmass}). To obtain the gravitino mass of the order of 
1 TeV, one has to adjust $\Lambda$ to be of the order of 
$10^{13}$ GeV when one constructs a realistic model by appropriately 
breaking the $E_8$ gauge group at the hidden wall.

In the minimization of the potential 
we have implicitly used the leading order approximation.
As was explained in a previous section, the next to leading order
correction gives the non--trivial dependence of the background metric
on $x^{11}$. Then the Einstein--Hilbert action in eleven dimensions
gives additional contribution to the scalar potential in the
four--dimensional effective theory, which shifts the vevs of the
$G_{IJKL}$. As a consequence, $F_S$ will no longer
vanish. Though this may be significant when we discuss soft masses,
it does not drastically change our estimate of the gravitino mass
(\ref{eq:gravitino-mass}) and our main conclusion drawn here is still
valid after the higher order corrections are taken into account.

%
%
\section{Soft supersymmetry breaking terms}

In the previous section, we have shown that the gaugino condensation
breaks supersymmetry both in the weakly coupled heterotic string and
in the heterotic $M$--theory. We chose $\Lambda$ in such a way that the
gravitino mass appeared in the TeV--range. In this section we shall 
discuss the soft supersymmetry breaking terms that appear in the 
low--energy effective theory as a consequence of this nonzero 
gravitino mass.

We first give the relevant formulae for gaugino and scalar masses in
the observable sector. Given the gauge kinetic function $f_6$ in the 
observable sector, the gaugino mass is calculated to be
\begin{equation}
     m_{1/2}=\frac{\partial f_6}{\partial \phi^i} 
             \frac{F^{i}}{2\mbox{Re} f_6},
\label{eq:gaugino-mass}
\end{equation}
where $\phi^i$ symbolically denote hidden sector fields responsible
for supersymmetry breakdown. 
Writing the K\"{a}hler potential 
\begin{equation}
     K= \hat K(\phi^i, \phi^{*}_i) 
          + Z(\phi^i, \phi^{*}_i) C^* C
          +(\mbox{higher orders in } C, C^*),
\end{equation}
one can also calculate the mass of a matter field $C$ \cite{KL,BIM} 
\begin{equation}
  m^2_0=m_{3/2}^2 
 -F^i F^*_j  \frac{Z_{i}^j-Z_i Z^{-1} Z^{j}}{Z}.
\label{eq:scalar-mass}
\end{equation}
Here a vanishing cosmological constant is assumed.

Using the classical approximation naively, these formulae lead to a
surprising result. All soft masses vanish. At the basis of this fact
it had been suggested that the gravitino mass could be arbitrarily
high, still leading to softly broken supersymmetry in the TeV
range. It has been observed meanwhile that this surprising result is
an artifact of the approximation and it is now commonly accepted that
generically the soft masses tend to be of the order of the gravitino
mass or at least not arbitrarily small compared to it. In general the
result for the soft scalar masses is strongly model dependent.  We
shall see in the following that the situation concerning the gaugino
mass is less model dependent but varies when we go from the weakly to
the strongly coupled case \cite{NOY}.

%
%
\subsection{Weak coupling case}

We start again with the weakly coupled case.  At the leading order 
(tree level), the gauge kinetic function for the observable sector 
is simply $f_6=S$, whereas the gaugino condensation gives $F_S=0$,
$F_T=m_{3/2}(T+ T^*)$. Thus, at this level, the gaugino mass vanishes. 
As was discussed in section 3, the gauge kinetic function receives 
corrections at one--loop order. Using eq. (\ref{eq:f6f8}), the 
gaugino mass is explicitly written as
\begin{equation}
   m_{1/2} =\frac{F_S + \epsilon F_T}{2 \mbox{Re}(S +\epsilon T)}.
\label{eq:gaugino-mass-weak}
\end{equation}
Note that $F_T/(T+ T^*) \sim m_{3/2}$. Also we expect $F_S$ to be
of the order of  $\epsilon T m_{3/2}$ due to the one--loop
corrections. Plugging them into the above expression, we obtain
\begin{equation}
     m_{1/2} \sim \frac{\epsilon T}{S} m_{3/2}.
\label{eq:gaugino-weak}
\end{equation}
Since in the weakly coupled case the ratio $\epsilon T/S$ is small,
the gaugino becomes much lighter than the gravitino.

Let us now consider the scalar masses. At the tree level, the
K\"ahler potential is 
\begin{equation}
    K=-\ln (S+ S^*) -3 \ln (T+ T^*) +(T+T^*)^n C^* C
   +(\mbox{higher orders in } C^* C),
\end{equation}
where $n$ denotes the modular weight of a field $C$. For a field with
$n=-1$ (untwisted sector in an orbifold construction), which
naturally appears in the simple truncation procedure, we recover the
previous formula (\ref{eq:G}). From eq.~(\ref{eq:scalar-mass}), it
follows that
\begin{equation}
   m_0^2=m_{3/2}^2 +\frac{|F_T|^2}{(T+T^*)^2}=(1+n)m_{3/2}^2.
\end{equation}
A scalar field with the modular weight $-1$ has a vanishing
supersymmetry--breaking mass at the leading order. 
It is an artifact of the approximation of reduction and
truncation (i.e. torus compactification) that the fields have
modular weight $-1$. A field whose modular weight is different from 
$-1$ has a mass comparable to the gravitino mass.  Though, as 
discussed in section 3, corrections at the one--loop level are model 
dependent, one expects they are of the order of 
$\epsilon T/ S m_{3/2}^2$. Summarizing these contributions, 
one obtains
\begin{equation}
   m_0^2=(1+n)m_{3/2}^2 + O( \frac{\epsilon T}{S} m_{3/2}^2),
\label{eq:scalar-weak}
\end{equation}
where the actual value of the second term depends on the model one
considers. A conclusion we can draw from eqs.\ (\ref{eq:gaugino-weak})
and (\ref{eq:scalar-weak}) is that the gaugino masses
tend to be  much smaller
than the scalar masses:
\begin{equation}
   m_{1/2} \ll m_{0} \leq O(m_{3/2}). \label{eq:relation-weak}
\end{equation}
Phenomenologically this relation might be problematic. Requiring 
that the gaugino masses are at the electro--weak scale,
eq.\ (\ref{eq:relation-weak}) 
would then imply that the masses of the squarks
and sleptons should be well above the 1 TeV region, which raises the
fine--tuning problem to reproduce the Fermi scale.  Another potential
problem is the relic abundance of the lightest superparticles (LSPs)
which are likely the lightest neutralinos in the present case. With the
parameters characterized by (\ref{eq:relation-weak}), the standard
computation of the relic abundances shows that too many LSPs 
 would (if stable) still be around today,
resulting in the overclosure of the Universe.

Thus in the weak coupling regime, one can conclude that, though the
gaugino condensation realizes the supersymmetry breaking, it tends to
lead to a picture where gaugino masses are generically smaller
than gravitino and scalar masses. A satisfactory situation might only
be achieved, if one fine--tunes the scalar masses in a way that they
become comparable to the gaugino masses.

%
%
\subsection{Strong coupling case}

Next we want to discuss how the situation changes when one
considers the strongly coupled case (heterotic $M$--theory). 

As in the weakly coupled heterotic string theory, the gaugino mass
vanishes at the leading order of the $\kappa^{2/3}$ expansions,
because $f_6={\sS}$ and $F_{\sS}=0$. Again the next to the 
leading order is important. The analogue of 
eq.\ (\ref{eq:gaugino-mass-weak}) in the strongly coupled case is 
\begin{equation}
  m_{1/2}=\frac{F_{\sS}
             +\alpha F_{\sT}}
              {2 \mbox{Re}({\sS}+\alpha {\sT})}.
\end{equation}
Thus we obtain, as before 
\begin{equation}
  m_{1/2} \sim \frac{\alpha {\sT }}{\sS} m_{3/2}.
\end{equation}
A crucial difference in this case, however, is the fact that the ratio
$\alpha {\sT }/ {\sS}$ is not a small number, but can be as
large as unity.  This is because the values of $\sS$ and $\sT$
inferred from our input variables (see section 2.2) suggests that we
are rather close to criticality (in which case the ratio becomes
unity). Thus we can conclude that, unlike the weakly coupled case, the
gaugino mass in the strongly coupled regime is comparable to the
gravitino mass.  This observation confirms the expectation that the
gravitino mass should be in the TeV--region and the gaugino
condensation scale $\Lambda \sim 10^{13}$ GeV. Because of the
simplicity of the mass formula (\ref{eq:gaugino-mass}) and the fact
that the gauge--kinetic function $f$ is stable in higher order
perturbation theory, the statement concerning the soft gaugino masses
is rather model independent.

The situation is more complicated in the case of the scalar masses 
which we consider now in the framework of heterotic $M$--theory. 
At the leading order we arrive at the same conclusions as in the weak 
coupling case, since the K\"ahler potential is identical in both cases.
In section 4, we calculated the corrections to the K\"ahler potential 
at the next to leading order, which reads
\begin{eqnarray}
   \hat K& =& -\ln ({\sS} +{\sS}^*) -3\ln ({\sT}+{\sT}^*)
\\
   Z &=& \frac{6}{{\sT}+{\sT}^*}
+\frac{2 \alpha }{{\sS} +{\sS}^*}
\end{eqnarray}
where the latter is valid for a field with the modular weight $-1$. 
Now using the formula (\ref{eq:scalar-mass}) one may be able to
calculate the scalar masses, with the result
\begin{eqnarray}
   m_{0}^2=m_{3/2}^2 
            -\frac{2-\frac{1}{1+\delta}}{1+\delta}
          \frac{|F_{\sT}|^2}{({\sT}+{\sT}^*)^2}
            -\frac{\delta(2-\delta \frac{1}{1+\delta})}{1+\delta}
          \frac{|F_{\sS}|^2}{({\sS}+{\sS}^*)^2}
 \nonumber \\
          -\frac{\delta}{(1+\delta)^2}(F_{\sS} F^*_{\sT}
           +F^*_{\sS} F_{\sT}) \label{eq:scalar-mass-strong}
\end{eqnarray}
where 
\begin{equation}
          \delta\equiv \frac{\alpha}{3}
      \frac{{\sT}+{\sT}^*}{{\sS}+{\sS}^*}.
\end{equation}
We can clearly see from this expression that the structure
obtained in the leading order is badly violated.   
Given the fact that the expansion parameter 
$\alpha({\sT}+{\sT}^*)/({\sS}+{\sS}^*)$ is 
of order unity it is no longer possible to fine tune the
scalar masses (by choosing modular weight $-1$ for all of them)
to a small value and then hope that the corrections respect this
fine tuning. In addition the scalar masses depend strongly on the
form of the K\"ahler potential which, in contrast to the
gauge kinetic function, receives further corrections in higher 
order. Thus detailed statements about the scalar masses are
very model dependent.

In summary we can, however, conclude 
with the qualitative statement that in the strong coupling regime,
\begin{equation}
   m_{1/2} \sim m_{0} \sim m_{3/2}. \label{eq:relation-strong}
\end{equation}
This contrasts with the
relation (\ref{eq:relation-weak}) for the weak coupling regime
and represents an important improvement concerning 
phenomenological applications. In the strongly coupled case,
the difference between dilaton-- and moduli--dominated supersymmetry
breakdown seems less pronounced than it is in the weakly coupled case.

%
%
\section{Summary and outlook}

We have presented a consistent framework of supersymmetry breaking
and soft breaking terms triggered by the gaugino condensate
at the hidden wall. In the strongly coupled case, in complete analogy 
to the weakly coupled case, the gravitino mass $m_{3/2}$ is related 
to the gaugino condensation scale $\Lambda$ as
\begin{equation}
   m_{3/2} \approx \frac{\Lambda^3}{M_{Planck}^2}.
\end{equation}
Furthermore, as explained in section 6, the soft masses are of
the order of the gravitino mass. This implies that these masses
should be in the TeV range in order to solve the naturalness problem 
of the Higgs boson mass in the supersymmetric framework. This 
requires that $\Lambda$ should be around $10^{13}$ GeV, three orders 
of magnitude smaller than the GUT scale (the compactification scale) 
and thus the 11D Planck scale as well. 
The gauge coupling constant at the $E_8$ wall, where the gaugino
condensate is supposed to occur, is larger than the  one at the
$E_6$ wall.  If the eleventh dimensional radius $\rho$ approaches
the critical radius $\rho_{crit}$, the $E_8$ gauge coupling
constant becomes strong at a scale as large as the GUT scale, 
and the running coupling constant will blow up at that scale
already. Then the gaugino condensation scale $\Lambda$, which is 
approximately identified with the blow--up energy scale, would become 
too large. For a value of $\Lambda \sim 10^{13}$ GeV, $\rho$ should 
(although close) not be too close to the critical value so that the
gauge coupling constant does not blow up immediately. This gives a
constraint on the constant $\alpha$ (defined in (\ref{eq:alpha})), 
which depends on the detailed properties of the Calabi--Yau manifold 
under consideration.  In any case it is probably necessary
to break  the hidden $E_8$ to a smaller group  to obtain a smaller
coefficient of the $\beta$--function.  These considerations
should be kept in mind when one attempts to 
construct a realistic model.

The fact that the gravitino mass cannot be arbitrarily large, but
should lie in the TeV range in the heterotic $M$--theory regime
suggests that the theory 
might share a problem already encountered in the weakly coupled
case \cite{PP,We,EKN}.  Late time decay of the gravitinos would
upset the success of the standard big--bang nucleosynthesis
scenario. This problem is rather universal in most of the supergravity
models where breakdown of supersymmetry is mediated through
gravity. Indeed this is not really a serious difficulty, 
but just implies that the universe underwent inflationary
expansion followed by reheating at a relatively low temperature 
($T < 10^9$ GeV for $m_{3/2}=1$ TeV \cite{KM}), in which the 
gravitino number density is diluted by the inflation and the low 
reheat temperature suppresses gravitino production after that.

A main difference between the weakly and the strongly
coupled case manifests itself when we consider
phenomenological issues associated with the soft
masses. In the weakly coupled string case, the gaugino condensation
scenario gives a very small gaugino mass compared to the scalar masses.
For a typical size of the compactification radius of the 6D manifold,
the gaugino mass is shown to be more than one order of magnitude
smaller than the scalar mass (see for example eqs. (7.20) and (7.24)
(with $\sin \theta \rightarrow 0$ limit) of ref. \cite{BIM} for more
detail).  This hierarchy among the soft masses obviously raises a
naturalness problem. With gaugino masses of the order of 100 GeV,
the scalar masses would be far above 1 TeV, requiring fine tuning to
obtain the electroweak symmetry breaking scale. This causes
problems for explicit model building.
Another phenomenological difficulty caused by the small gaugino mass
arises in the context of relic 
abundances of the lightest superparticles (LSPs).
Under the assumption of $R$--parity conservation, the LSP is stable
and remains today as a dark matter candidate. Given the superparticle
spectrum in the weak coupling regime, the bino, the superpartner of
the $U(1)_Y$ gauge boson, is most likely to be the LSP. To evaluate
the relic abundances of the bino, one has to know its annihilation
cross section (see ref.~\cite{JKG} and references therein). In our
case, the bino pair annihilates into fermion (quarks and leptons)
pairs via t--channel scalar (squarks and sleptons) exchange. The cross
section is roughly proportional to
\begin{equation}
   \sigma \propto \frac{m_{\tilde B}^2}{m_{\tilde f}^4}
\end{equation}
where $m_{\tilde B}$ is the bino mass and $m_{\tilde f}$ represents a
scalar mass. As the scalar becomes heavier, the cross section is
suppressed, yielding a larger relic abundance. Indeed when the scalar
mass is more than an order of magnitude larger than 
the gaugino mass, a standard
calculation shows that the relic abundance exceeds the critical value
of the universe. This
overclosure is a serious problem 
in the weakly coupled case.

In the strong coupling regime, the gaugino acquires a mass comparable
to the gravitino mass and the scalar masses. 
Thus the above two problems do not appear.
All the soft masses are in the same range. If this is not
far from the electroweak scale, one can naturally realize the
electroweak symmetry breaking at the correct scale without fine
tuning.  Moreover in this scenario, the annihilation cross section of
the bino becomes larger, and thus we can obtain a relic abundance
compatible with the observations.  In some  regions
of parameter space we may
even realize a situation where the LSP is the dominant component of
the dark matter of the universe.

A characteristic of the mechanism of gaugino
condensation is the fact that it is the $T$ field 
that plays the dominant role in the breakdown of supersymmetry.
In this scenario scalar fields 
with different modular weight will have different
masses, which may cause problems
with flavor changing neutral currents (FCNC). In
the strong coupling case, the situation may be improved 
through the presence of a large
gaugino mass which contributes to the scalar masses at low energies
through radiative corrections that can be computed via
renormalization group methods. In a situation where scalar masses
at the  GUT scale are small enough, this universal radiative 
contribution might wash out nonuniversalities and avoid problems
with FCNC. Details of the superparticle phenomenology
in the strongly coupled case, including the issues outlined above,
will be discussed elsewhere~\cite{KNOY}.

Eqs.\ (\ref{eq:f6f8}) (in the weak coupling case) and
(\ref{eq:f6f8alpha}) (in the strong coupling case) show that the
imaginary part of the complex scalar fields, $S$ and $T$, has an
axion--like coupling to the gluon fields.  In the weakly coupled case,
world--sheet instanton effects \cite{DSWW} and possibly other
non--perturbative effects give non--negligible contributions to the
potential. Then the axion candidates receive masses comparable to the
gravitino mass, and they do not solve the strong $CP$ problem.
However, in the strongly coupled case, it has been argued that these
non--perturbative contributions originated at high energy physics might
be suppressed to a negligible level \cite{BD,BD2,Choi}.  
If this is the case, a linear
combination of the Im$\sS$ and Im$\sT$ will play a role of the
axion, whose potential is dominated by the QCD contribution. Then this
axion, referred to as the $M$--theory axion, will be able to solve the
strong $CP$ problem. A word of caution should be added here, since
a reliable calculation of these world sheet nonperturbative effects
has only been performed in the weakly coupled case \cite{LMN}. The
above argumentation in the M--theory framework 
uses the implicit assumption that those
couplings remain as weak as in the case of the weakly coupled string,
an assumption that might not be necessarily correct.
Apart from that, the axion decay constant in this case becomes as
as large as $10^{16}$ GeV, which leads to the potential problem that 
the energy density of the coherent oscillation of the axion field
exceeds the critical energy density of the universe. This problem 
could be solved if the entropy production occurs after the QCD phase 
transition when the axion gets massive, or if this world is almost 
$CP$ conserving and the initial displacement of the axion field is 
very small.  The direct detection of the
relic axions with such a large decay constant would be extremely
difficult. However the $M$--theory axion may give a significant
contribution to the isocurvature density fluctuations during the
inflationary epoch, which may be detectable in future satellite
observations \cite{KY}. It remains to be seen whether this mechanism
leads to a satisfactory solution of the strong CP--problem.

In any case we have seen that the M--theoretic version of
the heterotic string shows some highly satisfactory 
phenomenological properties concerning the unification of
fundamental coupling constants as well as the nature 
of the soft supersymmetry breaking parameters.

\section*{Acknowledgments}

We would like to thank Jan Conrad for useful discussions. 
This work was supported by
the European Commission programs ERBFMRX--CT96--0045 and CT96--0090
and by a grant from Deutsche Forschungsgemeinschaft SFB--375--95.
The work of M.O. was partially supported by
the Polish State Committee for Scientific Research grant 2 P03B 040 12.
The work of M.Y. was partially supported by
the Grant--in--Aid for Scientific Research from the Ministry of
Education, Science and Culture of Japan No.\ 09640333.

\vfill\eject

%
%

\end{document}